\documentclass[twocolumn,showpacs,preprintnumbers,prl,nofootinbib]{revtex4-1}
\usepackage{graphicx,amsmath,amssymb,xcolor,hyperref}

\begin{document}

\preprint{CERN-PH-TH-2015-071}
\author{Florian Goertz}
\affiliation{Theory Division,\\
CERN, 1211 Geneva 23, Switzerland}
\email{florian.goertz@cern.ch}
\title{Electroweak Symmetry Breaking without the \boldmath{$\mu^2$} Term}

\date{\today}


\begin{abstract}

We demonstrate that from a low energy perspective a viable breaking of 
the electroweak symmetry, as present in nature, can be achieved
without the (negative sign) $\mu^2$ mass term in the Higgs potential,
thereby avoiding completely the appearance of relevant operators, featuring coefficients 
with a positive mass dimension, in the theory.
We show that such a setup is self consistent and not 
ruled out by Higgs physics. In particular, we point out that it is the 
lightness of the Higgs boson that allows for the electroweak symmetry
to be broken dynamically via operators of $D\geq 4$, consistent with
the power expansion. Beyond that, we entertain how
this scenario might even be preferred phenomenologically
compared to the ordinary mechanism of electroweak symmetry
breaking, as realized in the Standard Model, and argue that
it can be fully tested at the LHC.
In an appendix, we classify UV completions that could lead to such a setup, 
considering also the option of generating all scales dynamically.
\end{abstract}

\maketitle

\section{Introduction}
\vspace{-1.5mm}

We made important progress in understanding nature by uncovering its symmetries.
In particular, the very basis of the theories describing our universe at the most fundamental level, i.e. the Standard Model (SM) of particle physics and general relativity, are symmetries. However, some of these (local) symmetries are not manifest in the vacuum but rather broken spontaneously. This is reflected by the fact that the mediators of the weak force are massive, as are the (chiral) building blocks of matter, which is essential for the existence of the universe as we see it.
A common lore, in particular after the discovery of
the Higgs boson at the CERN Large Hadron Collider, 
is that this electroweak symmetry breaking (EWSB) is 
triggered by a negative-sign mass term
for a scalar Higgs-doublet field, either introduced by hand or 
generated dynamically. The final ultra-violet (UV) completion of
the SM is expected to generate (approximately) a Higgs potential 
of such a form.
However, as we will entertain in this article, a more fundamental theory of nature 
could also have an opposite low energy ($\gtrsim$ weak scale) limit, where the appearance of 
relevant operators, such as the negative Higgs-mass squared, is completely avoided. As we will show, in such 
a scenario an 'irrelevant' $D=6$ operator of the type ${\cal O}_6= |H|^6$ could 
induce a non-trivial vacuum for the scalar 
sector. %\footnote{Such operators are expected to be present on general grounds since the SM is not anticipated to be the final theory of nature.}

We will point out that it is the lightness of the Higgs boson $m_h^2\ll v^2$, that allows to consider this special setup, where the operators in the Higgs potential
are not only deformed in a sub-leading way, as a phenomenologically viable alternative 
to the SM form, fully trading the $(D=2)$ $\mu^2$-operator for ${\cal O}_6$
within a consistent effective field theory (EFT).
Entertaining the viability of this limit is of utmost importance as it clarifies the question if a mass term is required in the 
Higgs-boson potential in order to spontaneously break the symmetries that induce the forces of nature.

After having demonstrated the self-consistency of the setup, we will turn to the phenomenology of the model. 
The most important collider observable is Higgs-pair production, where we will show that the LHC is capable 
of fully testing the pure version of the idea. 
Beyond that, the model has intriguing consequences for cosmology. We will see that it just lies in the correct ballpark 
such as to allow for a strong first order phase transition, as required by electroweak baryogenesis.
While in the main part of the paper we just treat the $\mu=0$ Higgs potential as a distinct and interesting boundary condition 
that a potential UV model could fulfill, thereby opening up a new direction in model building, in the appendix we will present and classify 
possible ideas for UV completions.
\vspace{-1mm}

\section{The Form of the Higgs Potential}
\vspace{-1.5mm}

We consider the SM without relevant operators and instead augment the Higgs potential with a dimension-6 term 
$c_6/\Lambda^2\, {\cal O}_6$ such that it takes the simple form
	\begin{equation}
	\label{eq:pot}
	V(H)= \lambda |H|^4 + \frac{c_6}{\Lambda^2}  |H|^6\,,
	\end{equation}
where all dimensionfull parameters are either zero or at the cutoff of the theory. 

Inspecting the form of the potential, a first observation is that a stable and non-trivial minimum at $|H|^2>0$ 
should be possible if $\lambda < 0$ and $c_6 >0$. In the following, we will check if such a minimum is also viable 
phenomenologically. For any given cutoff scale $\Lambda$, we can first calculate the position of the minimum,
{\it i.e.}, the vacuum expectation value (vev), denoted as $\langle |H|^2\rangle \equiv v^2/2$, via $\partial V/ \partial 
|H|^2=0$.
We find 
	\begin{equation}
	\label{eq:v}
	v^2 = -\frac 4 3 \frac\lambda{c_6} \Lambda^2\,.
	\end{equation}
Clearly, the minimization condition only fixes the relative size of the coefficients $\lambda$ and $c_6/\Lambda^2$. %\footnote{We
%keep the cutoff scale $\Lambda$ explicit here, since it is related to a physical mass scale that we want to discuss later.}
In turn, an electroweak-scale vev can be obtained without the need for a large coefficient of the $D=6$ operator
${\cal O}_6$. The size of the latter will however get fixed by the mass of the physical Higgs excitation around the vev, $h$, 
where in unitary gauge  $H = 1/\sqrt 2 ( 0,  v + h)^T$. This is given as $m_h^2= \partial^2 V/ 
\partial h^2\large{|}_{h=0}$, leading to
	\begin{equation}
	\label{eq:mh}
	m_h^2 =3 v^2 \lambda  + \frac{15}{4} \frac{c_6}{\Lambda^2} v^4\,.
	\end{equation}
The consequences of these relations will be scrutinized in the next section.
\vspace{-1mm}

\section{Self Consistency of the Setup}
\vspace{-1.5mm}

\begin{figure}[!t]
	\begin{center}
	\includegraphics[height=1.6in]{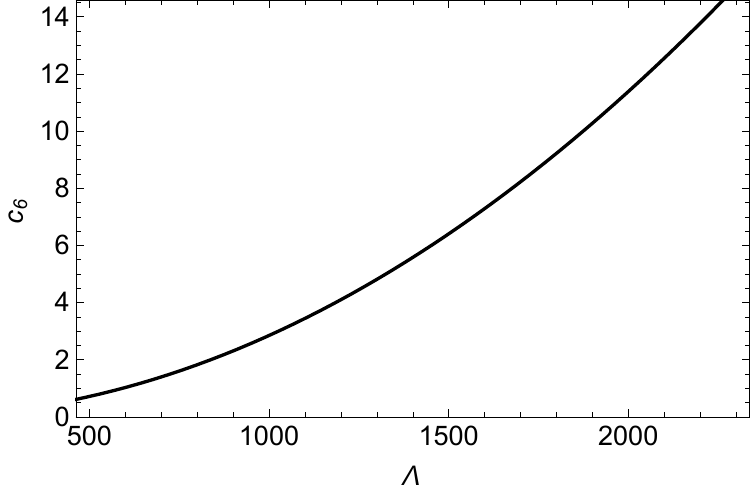}\\[-2mm]
	\caption{\label{fig:c6lam} Required value of the Wilson coefficient $c_6$ in dependence on the cutoff $\Lambda$. 
	See text for details.\\[-7mm]}
	\end{center}
	\end{figure}

We will now examine quantitatively, if it is possible to generate the vev and the Higgs mass at the
correct values in a self-consistent way with $\mu=0$, keeping the parameters in the range of the validity of the EFT.
To that extent, we first solve eqs.~(\ref{eq:v}) and (\ref{eq:mh})  for the two free parameters in the 
potential, $\lambda$ and $c_6$, expressing them in terms of the vev, fixed by the Fermi constant as 
$v=246$\,GeV, and the Higgs mass $m_h\approx125$\,GeV. We obtain 
	\begin{equation} 
	\lambda= - \frac{m_h^2}{2 v^2}\approx-0.13\,,\quad c_6= \frac{2 m_h^2}{3 v^2} 
	\frac{\Lambda^2}{v^2}\approx 2.8\, \frac{\Lambda^2}{{\rm TeV}^2}.
	\end{equation}
We inspect that, since $m_h^2/v^2\approx 1/4 \ll 1$, a large cutoff $\Lambda^2 \gg v^2$ is possible while
keeping $c_6 \sim {\cal O}(1)$. We can thus see explicitly that it is the lightness of the Higgs boson which allows for the mechanism to work naturally. 
% Indeed, eqs. (\ref{eq:v}) and (\ref{eq:mh}) taken together lead to $m_h^2 \ll v^2$
% for $\Lambda^2/c_6 \gg v^2$. 
The required $c_6$ versus the cutoff $\Lambda$ is visualized in Figure~\ref{fig:c6lam}.%, where $c_6 = 4\pi$ is given by the horizontal dashed line for illustration.
In particular, setting $\Lambda=0.8\,$TeV ($\Lambda=1$\,TeV) requires only $c_6=1.8$ 
($c_6=2.8$) while even $\Lambda=2\,$TeV is still possible in a rather strongly coupled setup 
with $c_6=11.4$.
On the other hand, around $\cal{O}({\rm several})$ TeV, at the latest, new physics (NP) would be expected.
If the new states are uncolored (which we will assume in the following), such mass scales clearly 
introduce no tension with 
current LHC limits. Moreover, we have checked that for all values of the cutoff considered above, the inclusion 
of a $D=8$ operator with ${\cal O}(1)$ coefficient alters the numerical results by only a few per cent 
or less.

We will now study more detailed the correlation between the needed size of the coefficient $c_6/\Lambda^2$ 
and the physical parameters in the Higgs sector %, $m_h$ and $v$, 
stressing that only a limited part of the larger parameter space, considered before the 
discovery of the Higgs boson, is viable in our model.
	\begin{figure}[!t]
	\begin{center}
	\includegraphics[height=1.62in]{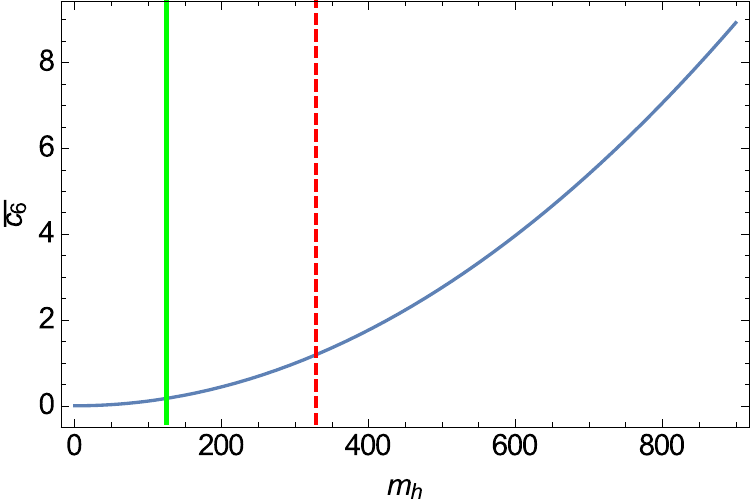}\\[-2.2mm]
	\caption{\label{fig:mh} Required $\bar c_6$ in dependence on $m_h$.\vspace{-4.6mm}}
	\end{center}
	\end{figure}

\begin{figure}[!t]
	\begin{center}
	\includegraphics[height=1.65in]{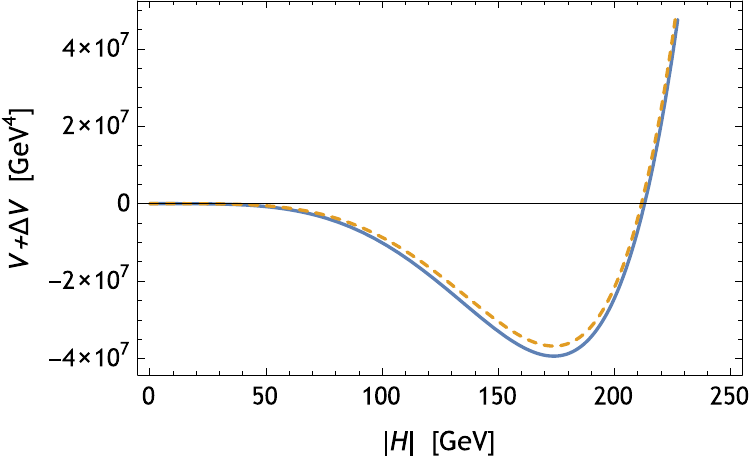}\\[-1.6mm]
	\caption{\label{fig:V} Blue curve: Higgs potential (\ref{eq:pot}), employing the physical
				$m_h$ and~$v$. Red dashed curve: Higgs potential, including the SM one-loop 
				corrections (\ref{eq:DV}), leading to~the~shifts~(\ref{eq:Dc}).\\[-7.9mm]}
	\end{center}
	\end{figure}

In Figure~\ref{fig:mh} we depict the required value of $\bar c_6 \equiv c_6 v^2/\Lambda^2$
(normalized to the weak scale) \mbox{versus} the Higgs-boson mass. 
While one can already estimate $\bar c_6 \sim 1$ as an upper bound on the viable parameter 
space of the perturbative EFT, this can be made more rigorous
by studying the limit following from (perturbative) unitarity, applying the optical theorem.
In fact, it is straightforward to show that unitarity in scalar-boson scattering in our 
model bounds \mbox{$|a_0^\infty|=7m_h^2/(8\pi v^2)\,<\,1/2$}, which is visualized by the red dashed line, 
corresponding to $|\bar c_6| \lesssim 1.2$.\footnote{This high energy constraint is approached 
quickly after the Higgs threshold, within the validity of the EFT considered here.} Thus, a heavy Higgs boson of only $m_h\gtrsim 300$\,GeV would have already 
basically invalidated our approach. The same is true for a vev of $v<100$\,GeV (keeping~$m_h=125$\,GeV). 
%which we show for completeness in the lower panel of Figure~\ref{fig:mh}. 
The experimental values \mbox{$m_h=125$\,GeV} and $v=246$\,GeV, visualized by a green vertical line, are however in 
perfect agreement with a reasonable value of $\bar c_6\approx0.17$. The potential~(\ref{eq:pot}), employing 
these values, is plotted as a solid blue line in Figure~\ref{fig:V}.
It exposes the expected mexican-hat form with a stable minimum at a non-trivial field value.
We conclude that, while it would have been easily possible that the numerical values of the mass
scales generated in nature after EWSB would have excluded our setup, the actual values just
lie in a range that allows for EWSB to be triggered by a single $D=6$ operator instead of a negative mass
squared term.

Finally, note that although within the low-energy theory discussed here the only physical (suppression) scale~$\Lambda$ can always be 
factored out of loop integrals and never enters dynamically, the question of the potential full
absence of the $\mu^2$ term beyond the tree level should be eventually addressed within a UV completion, providing a reason for its absence
(in the best of all cases avoiding tuning).
Accordingly, the peculiar setup itself does {\it not} provide a new solution to the hierarchy problem -
in fact the main focus of this work is to show its (non-trivial) phenomenological viability and special predictions.

To conclude this section, we show that the inclusion of the SM quantum corrections to the potential, generating a
term of the form $|H|^4 \log(H^2/\mu_r^2)$ \cite{Coleman:1973jx}, only corresponds to a small 
perturbation of our setup.
Neglecting the tiny impact of light quarks, the SM contributions to the one-loop Coleman-Weinberg potential 
are given by (see, {\it e.g.}, \cite{Quiros:1999jp})
	\begin{equation}
	\label{eq:DV}
	\Delta V =  \frac{1}{64 \pi^2} \sum_{i=W,Z,h,\chi,t} n_i\, M_i^4(H) 
	\left[ \log \frac{M_i^2(H)}{\mu_r^2} - C_i \right] \,.
	\end{equation}
Here, the tree-level field-dependent mass terms read
	\begin{equation}
	\begin{split}
	m_W^2(H)&=\frac{g^2}{2} H^2,\ m_Z^2(H)=\frac{g^2+{g^\prime}^2}{2} H^2,\\
	m_h^2(H)&=6 \lambda H^2,\ m_\chi^2(H)=2 \lambda H^2,\\[1mm]
	m_t^2(H)&=y_t^2 H^2\,,
	\end{split}
	\end{equation}
where we have dropped contributions suppressed by $\Lambda^2$,
the numbers of degrees of freedom are
	$
	n_W=6,\ n_Z=3,\ n_h=1,\ n_\chi=3,\ n_t=-12,
	$
and the constants $C_i$ are given by
	$
	C_W=C_Z=5/6,\ C_h=C_\chi=C_t=3/2\,.
	$ 
In the end, the top quark furnishes the dominant correction. 
Adding (\ref{eq:DV}) to (\ref{eq:pot}), setting the renormalization scale to $\mu_r=v/\sqrt 2$, and solving 
for $c_6$ and $\lambda$ that reproduce correctly $v$ and $m_h$, leads to the shifts
	\begin{equation}
	\label{eq:Dc}
	\Delta \lambda \approx -0.033,\,\quad \Delta \bar c_6 \approx 0.022\,,
	\end{equation}
which is a ${\cal O}(10\%)$ effect. We show the resulting potential as a red dashed line in 
Figure~\ref{fig:V}. It becomes a little bit flatter before the zero of the undisturbed potential and a 
bit steeper afterwards. Moreover, there arises a tiny maximum at low values of $|H|$, such that the origin is
a minimum - which however lies much higher than the global minimum at 
$|H|=v/\sqrt 2$.
\vspace{-1mm}

\section{Phenomenology}
\vspace{-1.85mm}
Beyond the potential direct discovery of new states around the TeV scale, our model offers distinct signatures in 
Higgs-pair production and cosmology that we want to discuss in the following. 

First of all, the sizable coefficient $\bar c_6 \approx 0.2$ leads to a notable change in the production cross section of 
Higgs pairs, since ${\cal O}_6$ contributes to the trilinear Higgs-self interaction after EWSB. In fact, it decreases the cross 
section by ~$\sim (60-70)\%$. This is in a range that should be possible to exclude at the LHC with 
a luminosity of ${\cal L} \gtrsim$~600\,fb$^{-1}$, see \cite{Goertz:2014qta}.\footnote{Note that $\bar c_6\approx 0.2$ 
corresponds to $c_6\approx 1.45$ in the conventions used to present the final results in \cite{Goertz:2014qta}.}

	\begin{figure}[!t]
	\begin{center}
	\includegraphics[height=1.685in]{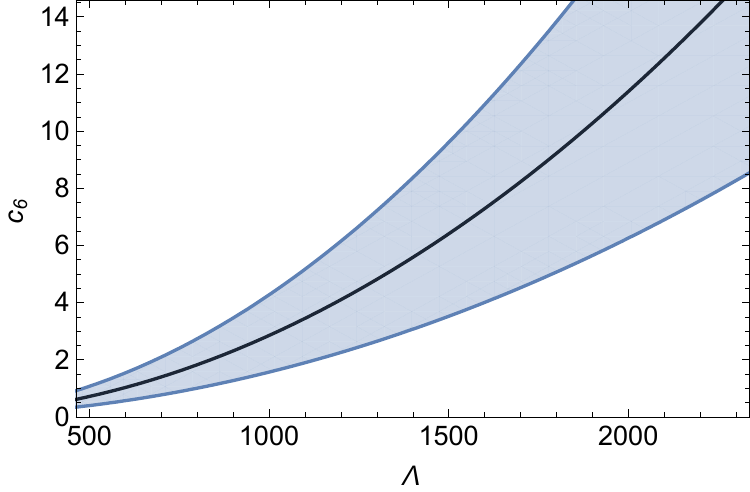}\\[-2.85mm]
	\caption{\label{fig:baryogen} The solid line depicts the required $c_6(\Lambda)$, 
	while the blue region allows for a first order electroweak phase transition triggering 
	electroweak baryogenesis.~See~text~for~details.\\[-7.45mm]}
	\end{center}
	\end{figure}
Beyond that, the presence of the operator ${\cal O}_6$ also modifies the electroweak phase transition. Without this 
operator, the phase transition is of second order for \mbox{$m_h=125$\, GeV} (see, {\it e.g.}, \cite{Quiros:1999jp}). This 
excludes the possibility of electroweak baryogenesis within the SM as there is no out-of equilibrium dynamics at the phase 
transition. On the other hand, a sizable contribution of ${\cal O}_6$  changes the Higgs potential such that a first order 
phase transition becomes possible for the physical Higgs mass~\cite{Grojean:2004xa}, allowing for electroweak baryogenesis
(if enough CP violation is present).
In Figure~\ref{fig:baryogen} we show again $c_6$ versus the cutoff $\Lambda$, where now the blue region corresponds
to a first order phase transition that leads to a stable $T=0$ minimum, while in addition sphaleron processes 
are sufficiently suppressed in the broken phase such as to not wash out the generated baryon asymmetry 
\cite{Grojean:2004xa}. The latter requirement leads to the condition $\langle h (T_c) \rangle /T_c \gtrsim 1$ at the critical 
temperature $T_c$.
Very interestingly, our $\mu^2=0$ solution just lies in the middle of the preferred region,
while the SM ({\it i.e.},~$c_6=0$) does not allow for electroweak baryogenesis.

We conclude that the required value of $\bar c_6$ leads to a very interesting phenomenology, allowing for pronounced
effects in Higgs-pair production as well as opening the possibility of the creation of our current universe via electroweak 
baryogenesis. This makes the setup avoiding relevant operators attractive on its own. Beyond that, it calls for an 
examination of how the effective potential (\ref{eq:pot}) could be generated - approximately or exactly - from a UV 
theory. This will be discussed in the Appendix.
\vspace{-1.2mm}

\section{Conclusions}
\vspace{-1.5mm}

As we know very little about the dynamics of EWSB or how the hierarchy problem is eventually solved in nature, various approaches
to EWSB should be examined and tested, in particular also from the low energy perspective, even if they might not
be the most obvious ones.
In this article, we have demonstrated that setting the notorious relevant operator $|H|^2$ in the Higgs potential to zero
and adding instead an operator ${\cal O}_6=|H|^6$ can lead to viable electroweak symmetry breaking,
thereby opening new directions in model building.
We pointed out that it is the lightness of the Higgs-boson that - perhaps unexpectedly -
 leads to this setup being self-consistent,
allowing a natural NP scale of
 $\Lambda \sim \mbox{$(1-2)$}$\,TeV. Eliminating the $\mu^2$ parameter and adding instead the $D=6$ coefficient $c_6$ keeps the theory
very predictive, since the number of parameters stays the same. 
In particular, the setup is fully testable in experiments currently under way, since relatively large changes in the Higgs-pair 
production cross section are predicted.

As it is a distinct theoretical limit, which also opens the possibility of generating 
our universe via baryogenesis at the weak scale and interestingly enough is not excluded by
Higgs phenomenology yet, the $\mu^2 \to 0$  model examined here should be considered as an alternative mechanism of 
breaking electroweak symmetry dynamically.
\vspace{1.5mm}

\acknowledgments
\paragraph{Acknowledgments}

\hspace{-3mm} I am grateful to Roberto Contino, Claude Duhr, Adam Falkowski, Gino Isidori, Andrey Katz, Valya Khoze, Matthew McCullough,
Andreas Weiler, and Jos{\'e} Zurita for useful discussions and comments.
My research is supported by a Marie Curie Intra European Fellowship within the 7th European Community Framework Programme 
(grant no.\ PIEF-GA-2013-628224).

\vspace{-3mm}
%\section{Appendix:}
\begin{appendix}
%\vspace{-3.5mm}
\section{APPENDIX:\, Possible UV Completions}
\label{sec:App}
\vspace{-2mm}

So far, the form of the Higgs potential (\ref{eq:pot}) was considered as a matching 
condition on the unspecified UV completion.
Now, we will discuss UV setups that could generate such a potential at the tree level. 
The general picture will be that the SM-like theory (featuring $\mu^2=0$) possesses no scale at the 
classical level and is then coupled to a sector that breaks scale invariance. 
%as depicted in Figure~\ref{fig:sectors}
Such an additional breaking is needed in the first place, since the breaking of scale invariance 
within the SM by dimensional transmutation is not sufficient to generate the Higgs mass of $m_h=125$\,GeV (see, {\it e.g.}, 
\cite{Hempfling:1996ht,Englert:2013gz}). 
The NP might itself respect scale invariance at the classical level, generating all masses dynamically.

%There, we will in particular provide 
%the connection to models that address the hierarchy problem by invoking the notion of classical scale invariance (CSI)
%\cite{Bardeen:1995kv,Meissner:2006zh,Meissner:2007xv,Englert:2013gz}, showing how in our model all scales could 
%be generated dynamically, in an orthogonal way compared to the usual models of CSI (see, {\it e.g.}, \cite{Hempfling:1996ht, 
%Chang:2007ki,Englert:2013gz} and references therein).
%
%	\begin{figure}[!t]
%	\begin{center}
%	\raisebox{-2mm}{\includegraphics[height=1.2in]{CSI.pdf}}
%	\caption{\label{fig:sectors} Pictorial representation of the breaking of scale invariance via interactions with
%	a new physics sector.}
%	\end{center}
%	\end{figure}

Let us however stress a difference compared to the usual approach,
often used in models employing classical scale invariance (CSI) as a building principle. In the latter, the $\mu^2$ term is 
forbidden at the tree level, but then regenerated spontaneously, usually via the (loop-induced) vev of an additional scalar 
singlet in a Higgs portal term, mimicking the usual SM Higgs potential  (see, {\it e.g.}, \cite{Hempfling:1996ht,Englert:2013gz},
as well as \cite{multi1,multi2} on general models that generate all scales dynamically).
In our approach, however, no relevant operator needs to 
be generated in the electroweak-scale theory at all. The breaking of scale invariance is induced in an orthogonal 
-\,possibly also spontaneous/dynamical\,- way, via an {\it irrelevant} operator, introduced by integrating out a heavy field 
that couples to the SM. This leads to a distinct low energy phenomenology and full testability of our 
setup. It provides a new minimal way of allowing for viable EWSB in the presence of the scale-invariant tree-level 
SM Lagrangian, that interestingly features $m_h \to 0$ in the decoupling limit $\Lambda \to \infty$.
\footnote{Also a combination, generating a very small (potentially even positive) $\mu^2$, while assisting with ${\cal O}_6$ 
to trigger EWSB in a theory respecting scale invariance at the tree level, might be interesting.} 

\begin{figure}[!t]
	\begin{center}
	\includegraphics[height=0.85in]{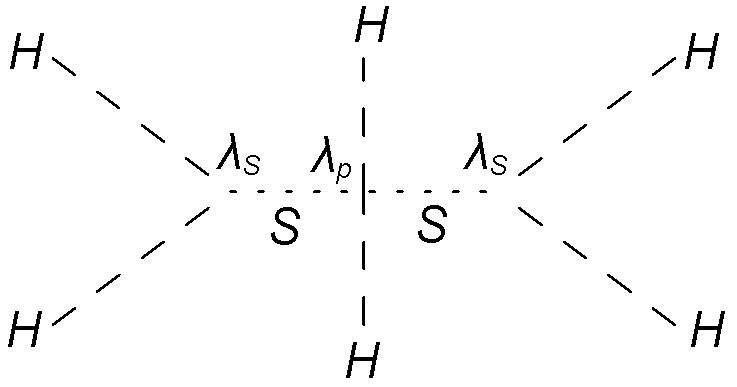}\\[-1.5mm]
	\caption{\label{fig:intout} Generation of ${\cal O}_6$ by integrating out the 
	singlet $S$.}
	\end{center}
	\end{figure}

We consider a scalar field $S$, singlet of $SU(2)_L$ with mass $M_S$, to 
generate the operator ${\cal O}_6$ at the tree level via the interactions
$M_S \lambda_S S |H|^2$ and $\lambda_p S^2 |H|^2$, see Figure~\ref{fig:intout},
leading to $c_6/\Lambda^2 \sim \lambda_p|\lambda_S|^2 /M_S^2$.
This allows the NP to be not too light, while a 
potential contribution to the $|H|^2$ operator could be deferred to the loop level (or beyond, in the 
presence of additional structures). To generate all scales dynamically, 
the dimensionfull coefficient 
in front of $\lambda_S$ could be thought of as a vev of a new field, or arise from a compositeness scale,
see below. Since at tree level only ${\cal O}_6$ is generated, one could entertain the possibility that
quadratically cutoff-dependent quantum corrections to $\mu^2$ are canceled in an extended NP sector,
e.g. by invoking (partial) supersymmetry or a twin Higgs mechanism,
such as to approximate, or even fully satisfy, (\ref{eq:pot}) in the full quantum theory. 
A related discussion on a complete cancellation of UV effects on the $|H|^2$ operator - 
which however there is generated again spontaneously - 
is given in \cite{Englert:2013gz} (see also \cite{Hempfling:1996ht}). 
Alternatively, the interaction terms might be cut off by a rapidly vanishing form factor
$\lambda_i(p)$, taming loops but not preventing a sizable tree 
level contribution to ${\cal O}_6$ via integrating out $S$ at zero momentum. 
Finally, one could just set the (renormalized) relevant operator to zero
at the matching scale. In any case, the scalar $S$ allows to entertain UV completions where 
EWSB could be driven by ${\cal O}_6$ and not by a negative sign $\mu^2$ term.
If $\langle S \rangle=0$ and $M_S \gg v$, with a significant fraction
not stemming from the SM-Higgs sector, it should also be save from current limits. 
%A survey of the impact of the new fields from the UV 
%completion on Higgs phenomenology (and precision observables) is left for future work. 
For an overview of constraints on scalar extensions of the SM see, {\it e.g.}, 
\cite{deBlas:2014mba}. Finally, extension of the vector-boson sector could also induce ${\cal O}_6$ 
at the tree level,~see~\cite{delAguila:2010mx}.

Note that all scales in the UV completion could be generated dynamically,
avoiding relevant operators not only in the IR limit $E\sim v$ but also in the 
shortest-distance UV theory, via considering the heavy fields to be composites
of a new strong interaction,
%with a condensation scale well above 1\,TeV to form a massive scalar singlet of $SU(2)_L$.
%These fields could be massless before they condense - 
%however the UV constituents could, depending on their quantum numbers, possibly also get a mass from the Higgs 
%sector after EWSB. 
%Beyond the mass of the composite state, as discussed above, one needs a scale to be present in the
%$S |H|^2$ interaction. This in turn could also be generated from the composite sector or arise 
%via the vev of a further singlet from the Coleman-Weinberg mechanism (see below).
%Alternatively, a scale could be generated 
or via the Coleman-Weinberg mechanism.
For the latter, a further scalar singlet could obtain a dynamical vev as explained before (see also \cite{Englert:2013gz}), 
inducing the mass of $S$ via a portal interaction (while direct portals to the SM Higgs could be suppressed
{\it e.g.} via geometrical sequestering \cite{ArkaniHamed:1999dc}).

Finally, nature might have chosen a  completely different way to generate ${\cal O}_6$, while avoiding the 
$\mu^2$ term, still to be found. A further analysis of the potential UV completions, including the examination 
of dark matter candidates, will be deferred to future work.

\end{appendix}

\end{document}